\documentclass[pre,aps,twocolumn,superscriptaddress,showpacs]{revtex4}
\newcommand{\beq}{\begin{equation}}
\newcommand{\eeq}{\end{equation}}
\newcommand{\bqn}{\begin{eqnarray}}
\newcommand{\eqn}{\end{eqnarray}}
\newcommand{\bqns}{\begin{eqnarray*}}
\newcommand{\eqns}{\end{eqnarray*}}
\newcommand{\bary}{\begin{array}}
\newcommand{\eary}{\end{array}}
\newcommand{\non}{\nonumber}
\usepackage{graphicx}%
\usepackage{dcolumn}
\usepackage{amsmath}
\begin{document}
\title{Unified treatment for classical waves in two-dimensional
media}
\author{Pi-Gang Luan}
\affiliation{Institute of Optical Sciences, National Central
University, Chung-Li 32054, Taiwan, Republic of China }
\author{Tzong-Jer Yang}
\affiliation{Department of Electrophysics, National Chiao Tung University,
Hsinchu 30050, Taiwan, Republic of China}
\date{\today}
\begin{abstract}
A unified treatment for the propagation of classical waves
in inhhomogeneous media is proposed.
We deal with four kinds of waves, they are
the acoustic wave in fluid, the elastic shear wave in two diemnsional solid,
and the E- and H-polarized electromagnetic waves in two diemnsional lossless
medium. We first show that a universal wave equation governing the wave motion
of all these four kinds of waves can be derived.
We then introduce an auxiliary field, and give the universal expressions for
the energy densities and energy flows of these waves.
This unified treatment provides intuitive insights, which may be helpful in
understanding the essential physics of various wave phenomena, or useful for
designing new photonic and sonic devices.
\end{abstract}
 \maketitle
\section{introduction}

Periodic media such as the photonic crystals \cite{Yab1,John1,Sakoda,Joan,Sig,CSK1,Zhang1}
and phononic crystals \cite{Kush1,Caba,Tor,Eco,Psa,Shengsci,Lambin,Gof,CSK2,Zhang2,Zhang3}
have received considerable attention in the past decades because of their interesting
physical properties and application potential in designing various new devices.
Waves propagating inside these media will be modulated by the periodic
structures. Therefore, the behavior of these waves will no longer be the same as they
do in a free space, and the so called frequency band structures appear \cite{Kittel}.
Besides, if we destroy the periodicity to make these structures
disordered, then under certain conditions there might appear another interesting phenomenon
called {\it Anderson localization} \cite{Ish, Shengedt,Zhen,Luan,Sampaio}, which is caused
by the multiple scattering of waves in the media.
Recently, the fascinating and unusual {\it negative refraction} (NR)
phenomenon \cite{Ves,Pendry} in a kind of artificial media called {\it left-handed materials}
(LHM) \cite{Smith1,Smith2} received even more attention and has already stimulated a large amount
of investagations \cite{ZH,Notomi,tHooft,JMW,VWV,Garcia,Ye,SSP,Wu,FES,FS,LJJ}.
Even more interestingly, the newest research results have shown that the NR-like phenomena can also happen in acoustic systems \cite{XZhang,Ye1}.

All of these developments are about classical waves \cite{Soklis,Soklis1}, including
the electromagnetic (EM) waves, the acoustic (AC) waves, and the elastic (EL)
waves. Clearly there are similarities between the EM and AC systems
(hereafter both the AC and EL systems will be denoted as the AC systems),
and in the past decade these similarities indeed stimulated some parallel developments about the studies
on EM and AC systems, such like the photonic and sonic crystals. Although these similarities have
been noticed since as early as in the 19th century \cite{Brln}, however,
up to now it seems that a systematic study and a complete understanding on them are
lacking. This situation leads to some inconveniences. For
example, there have been proposed useful guiding rules for designing photonic crystals \cite{Joan,Sig,CSK1,Zhang1}
and phononic crystals \cite{Gof,CSK2,Zhang2,Zhang3} with large band gaps. However,
it is not easy to understand the explicit meanings of
these rules. Besides, since in most studies the researchers put restrictions on the
values of the material parameters (for example, in a photonic
crystal one usually assumes the permeability $\mu=1$), the
usefulness of these rules might be very restrictive. By
extending the value of material parameters to the unexplored regions,
some interesting or unexpected phenomena might be found \cite{Zhen,Luan,Sampaio}.
An abstract and unified treatment are thus useful and can avoid unnecessary
repeated works. It can also help us to find the essential physics that
governs various wave phenomena. Furthermore, since this treatment
unifies the AC and EM systems, anyone who is familiar with only one side, say, the AC
systems, can ``translate" his knowledge to the other
side, and thus can help him to understand the apparently different systems more easily.

In this paper, we shall show that for AC wave
in fluid, the EL shear (ELSH) wave in two-dimensional solid, and the E- and
H-polarized EM waves in two-dimensional nonabsorptive media, a unified
theory can be constructed, which gives us a universal description of the dynamical
behaviors of the above mentioned classical waves. We believe that a further development
of this theory can help us much in clarifying the essential physics of the
classical wave systems.

This paper is organized as follows. The explicit wave equations
for the four kinds of waves mentioned before will be given in the
next section. A unified treatment for these waves is developed in
section III, in which we introduce an auxiliary field. Using
this auxiliary field, not only a universal wave equation, but also all physical
quantities such as energy densities and energy flows can be expressed in a
unified manner. We then discuss in section IV the the applications of this unified
theory. Section V is the
summary of this paper. Finally in Appendix A we derive
the conservation laws for elastic waves.

\section{The Wave equations}

In this section we first review the wave equations of the EL, AC
and EM waves, respectively. We then demonstrate that for
2D propagation, the wave equations for the above mentioned four kinds of
classical waves have the same form and thus can be treated in a
unified manner.

\subsection{Elastic waves}

Let ${\mathbf u}$ be the deformation of the medium at certain
point from its equilibrium position due to the stress. In terms of
Cartesian coordinate system, the propagation of EL waves in an
inhomogeneous  medium is governed by \bqn \rho\frac{\partial^2
u_i}{\partial t^2} &=&\frac{\partial}{\partial x_{i}}
\left(\lambda_e \frac{\partial u_j}{\partial x_j}
\right)+\frac{\partial}{\partial x_j}
\left[\mu_e\left(\frac{\partial u_i}{\partial x_j} +\frac{\partial
u_j}{\partial x_i}\right)\right]\non\\ &=&\frac{\partial
T_{ij}}{\partial x_j}, \label{elast} \eqn where $\rho$ is the mass
density, $\lambda_e$ and $\mu_e$ are the Lam\'{e} coefficients,
and \beq
 T_{ij}=\lambda_e\; (\partial_{l} u_{l})\delta_{ij}+\mu_e
\left(\partial_{i} u_{j}+ \partial_{j}u_{i}\right), \eeq is the
$ij$th component of stress tensor, all quantities are position
dependent. In terms of vector notations, the above
equation can also be written as \beq \rho \frac{\partial^2
u_{i}}{\partial t^2}= \frac{\partial}{\partial x_i}\left(
\lambda_e \nabla\cdot {\mathbf u}\right)+ \nabla\cdot (\mu_e
\nabla u_i + \mu_e \frac{\partial {\mathbf u}}{\partial x_i}).
\eeq In a homogeneous medium ${\mathbf u}$ can always be written
as the sum of a gradient and a curl, i.e., \beq {\mathbf
u}={\mathbf u}_l+{\mathbf u}_t =\nabla \Phi+\nabla\times
\vec{\Psi}, \eeq where ${\mathbf u}_l=\nabla\Phi$ and ${\mathbf
u}_t=\nabla\times\vec{\Psi}$ represent the longitudinal and
transverse components of the waves, respectively. The fields
$\Phi$ and $\vec{\Psi}$ satisfy \beq \nabla^2
\Phi-\frac{1}{c_l^2}\frac{\partial^2 \Phi}{\partial t^2}=0,
\;\;\;\;\;\;\nabla^2 \vec{\Psi}-\frac{1}{c_t^2}\frac{\partial^2
\vec{\Psi}}{\partial t^2}=0, \eeq where \beq c_l
=\sqrt{\frac{\lambda_e+2\mu_e}{\rho}},\;\;\;\;\;\;\;
c_t=\sqrt{\frac{\mu_e}{\rho}} \eeq are the phase velocities of the
longitudinal and transverse waves, respectively. From these
results the Lam$\acute{e}$ coefficients can be expressed as \beq
\lambda_e=\rho(c^2_l-2c^2_t),\;\;\;\;\;\; \mu_e=\rho c^2_t. \eeq

\subsubsection{Acoustic waves}

One special case of Eq.~(\ref{elast}) is the wave equation for AC
waves in fluid. In fluid, we assume $\mu_e=0$ (no shear
force) and thus $c_t=0$. Now we denote the phase velocity of the
wave as $c$, where $c=c_l$. The pressure $p$ and vibration
velocity ${\mathbf v}$ can be derived via the relations \beq
p=-\lambda_e \nabla\cdot {\mathbf u},\;\;\;\;\;\; {\mathbf
v}=\frac{\partial\mathbf u}{\partial t}.\label{pu} \eeq Substitute
(\ref{pu}) into (\ref{elast}) and one finds the wave equation \beq
\nabla\cdot\left(\frac{\nabla p}{\rho}\right) =\frac{1}{\rho
c^2}\frac{\partial^2 p}{\partial t^2},\label{we} \eeq and the
equation relating ${\mathbf v}$ to $p$:
\beq
 \frac{\partial
 {\mathbf v}}{\partial t} =-\frac{1}{\rho}\nabla p.
\eeq

\subsubsection{Elastic SH waves}

Another special case of Eq.~(\ref{elast}) is the wave equation for
ELSH waves. Suppose the medium considered is homogeneous along
the direction of $\hat{\mathbf z}$, and wave is propagating along
a path lying on the xy plane, then the ${\mathbf u}$ vector can always be
written as a sum of two decoupled vectors: a ${\mathbf u}_{xy}$
vector that lies on the xy plane, and a ${\mathbf u}_z$ vector that parallels
the z-axis. Now if ${\mathbf
u}_{xy}=0$, i.e., SH wave, then ${\mathbf u}=u \hat{\mathbf z}$
and the wave equation becomes \beq \rho \frac{\partial^2
u}{\partial t^2}= \nabla\cdot (\mu_e \nabla u)=\nabla\cdot (\rho
c^2_t \nabla u).\label{uu} \eeq For the convenience of the following
discussions, we define \beq \eta\equiv\frac{1}{\mu_e}=\frac{1}{\rho
c^2_t}, \eeq and thus Eq. (\ref{uu}) can be rewritten as \beq
\nabla\cdot \left(\frac{\nabla u}{\eta}\right)=\frac{1}{\eta c^2_t}
\frac{\partial^2 u}{\partial t^2}\label{uu1}. \eeq

\subsection{Electromagnetic waves}

Now we turn to the discussion of EM waves. Adopting the Gaussian
unit, and denoting the speed of light in vacuum as $c_0$, Maxwell's
equations in a source free and locally isotropic medium are written as
\beq
 \nabla\times {\mathbf E}=-\frac{1}{c_0}\frac{\partial
 {\mathbf B}}{\partial t} =-\frac{\mu}{c_0}\frac{\partial {\mathbf
 H}}{\partial t} \label{curle} \eeq \beq \nabla\times {\mathbf
 H}=-\frac{1}{c_0}\frac{\partial {\mathbf D}}{\partial t}
 =\frac{\epsilon}{c_0}\frac{\partial {\mathbf E}}{\partial t}
 \label{curlh}
\eeq and
\beq
 \nabla\cdot {\mathbf D}=\nabla\cdot({\epsilon\mathbf
 E})=0,\;\;\;\;
 \nabla\cdot {\mathbf B}=\nabla\cdot ({\mu\mathbf H})=0.
\eeq Here $\epsilon=\epsilon(\mathbf r)$ and $\mu=\mu(\mathbf r)$
are position dependent permittivity and permeability. From (\ref{curle}) and
(\ref{curlh}), we can esaily derive the wave equations in terms of
${\mathbf E}$ and ${\mathbf H}$ fields: \beq \nabla\times
\left(\frac{\nabla\times {\mathbf E}}{\mu}\right)
=-\frac{\epsilon}{c_0^2} \frac{\partial^2\mathbf E}{\partial
t^2}\label{ee} \eeq \beq \nabla\times \left(\frac{\nabla\times
{\mathbf H}}{\epsilon}\right) =-\frac{\mu}{c_0^2}
\frac{\partial^2\mathbf H}{\partial t^2}.\label{hh} \eeq

\subsubsection{E-polarized and H-polarized EM waves}

Like that in the discussion of the elastic waves, we now
assume that the medium is translational invariant along
$\hat{\mathbf z}$. It is well known that in this
case if an EM wave propagating along a path on XY plane,
then electromagnetic fields can always be decoupled into the
E-polarized and H-polarized waves.
For E-polarized waves, we mean ${\mathbf E}=E\hat{\mathbf z}$.
Remember that in this system
${\partial}(\mbox{anything})/{\partial z}=0$, thus we have
\beq
\nabla\times \left(\frac{\nabla\times {\mathbf E}}{\mu}\right)
=-\hat{\mathbf z}\nabla\cdot \left(\frac{\nabla E}{\mu}\right).
\eeq
So, Eq. (\ref{ee}) becomes
\beq
\nabla\cdot\left(\frac{\nabla E}{\mu}\right)
=\frac{\epsilon}{c_0^2} \frac{\partial^2 E}{\partial t^2}
=\frac{1}{\mu c^2} \frac{\partial^2 E}{\partial t^2}.\label{Ep}
\eeq
Similarly, for H-polarized waves, we have
\beq
\nabla\cdot\left(\frac{\nabla H}{\epsilon}\right)
=\frac{\mu}{c_0^2} \frac{\partial^2 H}{\partial t^2}
=\frac{1}{\epsilon c^2} \frac{\partial^2 H}{\partial t^2}.\label{Hp}
\eeq
Here $c=c_0/\sqrt{\epsilon\mu}$ represents
the speed of light in the medium.

\section{Unified treatment for classical waves in 2D systems}

All the wave equations of the four cases discussed
before have the same form
\beq
\nabla\cdot\left(\frac{\nabla U}{\alpha}\right)
=\frac{1}{\alpha c^2} \frac{\partial^2 U}{\partial t^2},\label{unify}
\eeq
thus their wave equations can be treated in a unified manner.
However, this is not enough if we want to construct a complete unified description.
Such a theory must also deal with the quantities like energy density and energy flow in a
unified manner. To achieve this goal, we now introduce a field whcih we call the
{\it auxilliary field}. We will show in the later discussion that the complete unified
description can be constructed using the wave equation of the auxilliary field.

Consider an auxiliary field satisfying the
wave equation
\beq
\nabla\cdot\left(\frac{\nabla \Phi}{\alpha}\right)-
\frac{1}{\alpha c^2}\frac{\partial^2 \Phi}{\partial t^2}=0,\label{awe}
\eeq
which has the same form as in Eq.~(\ref{unify}) with
$\alpha=\alpha(\mathbf r)$ and $c=c(\mathbf r)$ being
position-dependent parameters. Throughout the
paper we assume $\partial \Phi/\partial z=0$, i.e., ${\Phi}={\Phi}(x,y,t)$.
Under this assumption the system becomes effectly two
dimensional.
We define the
dynamical variables as $\varphi$ and ${\mathbf Q}$:
\beq
\varphi\equiv\frac{\partial\Phi}{\partial t},\;\;\;\;\;\;\;
{\mathbf Q}\equiv-\frac{1}{\alpha}\nabla \Phi.\label{def}
\eeq
We will see in the later discussion that $\Phi$ plays the role
of the {\it vector potential} in the electromagnetics.
Using the difinition (\ref{def}) the wave equation (\ref{awe}) can
be replaced by
\beq
\frac{\partial{\mathbf Q}}{\partial t}
=-\frac{1}{\alpha}\nabla\varphi,\;\;\;\;\;\;\;
\frac{\partial \varphi}{\partial t}
=-\alpha c^2\nabla\cdot {\mathbf Q}.\label{qphiwe}
\eeq
Note that if we choose the Lagrangian density
\beq
{\mathcal L}=\frac{1}{2\alpha c^2}
\left(\frac{\partial \Phi}{\partial t}\right)^2
-\frac{1}{2\alpha}\left(\nabla\Phi\right)^2
=\frac{1}{2\alpha c^2}\varphi^2
-\frac{\alpha}{2}{\mathbf Q}^2,
\eeq
then Eq.~(\ref{awe}) and (\ref{qphiwe}) can be deduced from the
Euler equation of motiom
\beq
\frac{\partial \mathcal{L}}{\partial \Phi}
-\frac{\partial}{\partial t}
\frac{\partial\mathcal{L}}{\partial \dot{\Phi}}
-\nabla\cdot
\left[\frac{\partial\mathcal{L}}{\partial(\nabla \Phi)}\right]=0.
\eeq
Here we have employed the notation
$\dot{\Phi}\equiv \partial\Phi/\partial t$.
With this lagrangian, the canonical momentum $\Pi$ and
the Hamiltonian ${\mathcal{H}}$ are given by
\beq
\Pi=\frac{\partial \mathcal{L}}{\partial \dot{\Phi}}
=\frac{1}{\alpha c^2}\dot{\Phi}=\frac{1}{\alpha c^2}\varphi.
\eeq
and
\bqn
{\mathcal{H}}&=&\Pi\dot{\Phi}-{\mathcal{L}}\non\\
&=&\frac{\alpha c^2}{2}\Pi^2
+\frac{1}{2\alpha}(\nabla \Phi)^2\non\\
&=&\frac{\alpha}{2}{\mathbf Q}^2+\frac{1}{2\alpha c^2}\varphi^2
\label{ham}.
\eqn

\subsection{The Energy conservation law}

In terms of Eq.~(\ref{awe}), we now derive the energy conservation
law. Multiplying $-\dot{\Phi}$ on Eq.~(\ref{awe}), we have
\[
-\nabla\cdot \left(\frac{\dot{\Phi}\nabla\Phi}{\alpha}\right)
+\frac{\nabla \dot{\Phi}\cdot\nabla \Phi}{\alpha}+\frac{\dot{\Phi}}{\alpha c^2}
\frac{\partial \dot{\Phi}}{\partial t}=0
\]
or
\beq
\nabla\cdot\left(-\dot{\Phi}\frac{\nabla\Phi}{\alpha}\right)
+\frac{\partial}{\partial t}
\left[\frac{1}{2\alpha c^2}{\dot{\Phi}}^2
+\frac{1}{2\alpha} (\nabla \Phi)^2 \right]=0.\label{consv}
\eeq
Define the energy flux ${\mathbf J}$ and energy density
$\mathcal{U}$ as
\beq
{\mathbf J}\equiv-\frac{1}{\alpha}\dot{\Phi}\nabla\Phi=\varphi\,{\mathbf Q}
\label{flux}
\eeq
and
\beq
{\mathcal{U}}\equiv\frac{1}{2\alpha}(\nabla \Phi)^2
+\frac{1}{2\alpha c^2}{\dot{\Phi}}^2
=\frac{\alpha}{2}{\mathbf Q}^2+\frac{1}{2\alpha c^2}\varphi^2,
\label{ene}
\eeq
then Eq.~(\ref{consv}) becomes
\beq
\nabla\cdot {\mathbf J}+\frac{\partial \mathcal{U}}{\partial t}=0,
\eeq
which is the energy conservation law (the``continuity equation" for the energy).
Comparing Eq.~(\ref{ene}) with (\ref{ham}) confirms that $\mathcal{U}$ is indeed
the energy density.

\subsection{Identification of ${\mathbf J}$ and ${\cal U}$
with known physical quantities}

\subsubsection{${\mathbf J}$ and ${\mathcal{U}}$ for acoustic waves in
fluid}

For the acoustic wave propagation in 2D we define
\beq
\Phi\equiv\int^t  p\,dt,\;\;\;\;\alpha\equiv\rho,
\eeq
then
\beq
\varphi= p,\;\;\;\;\;{\mathbf Q}={\mathbf v}.
\eeq
The energy flux ${\mathbf J}$ and energy density ${\mathcal{U}}$
are given by
\beq
{\mathbf J}=p\,{\mathbf v},\;\;\;\;\;\;\;
{\mathcal{U}}=\frac{1}{2}\rho\,{v}^2+\frac{1}{2\rho c^2}\,p^2,
\eeq
which are consistent with the results from the 3D formula (See appendix A).

\subsubsection{${\mathbf J}$ and ${\mathcal{U}}$ for elastic SH
waves in solid}

For the SH waves ${\mathbf u}=u(x,y,t)\,\hat{\mathbf z}$.
We define
\beq
\Phi\equiv u,\;\;\;\;\alpha\equiv\frac{1}{\mu_e}=\frac{1}{\rho c^2_t},
\eeq
then
\beq
\varphi=\dot{u}=v,\;\;\;\;\;
{\mathbf Q}=-\mu_e\nabla u=-{\mathbf T}\cdot\hat{\mathbf z},
\eeq
where ${\mathbf v}=v\hat{\mathbf z}$  represents
the vibration velocity of the media and ${\mathbf T}$ stands for the stress tensor.

Now we denote $\hat{\bf n}$ ($\|\nabla u$) as the direction of
${\mathbf Q}$, i.e.,
\beq
{\mathbf Q}=Q\,\hat{\mathbf n},\;\;\;\;\;Q=|{\mathbf Q}|.
\eeq
Suppose there is a region enclosed by a surface with normal
vector $\hat{\mathbf n}$ (here we have $\hat{\bf n}\bot \hat{\bf z}$). The surface exerts
a shear force ${\mathbf f}_s$
(along $\hat{\mathbf z}$.) on its
surrounding medium, which is given by
\beq
{\mathbf f}_s=-{\mathbf T}\cdot\hat{\mathbf n}=Q\,\hat{\mathbf z}.
\eeq
From this observation we find that ${\mathbf Q}$ in this
case is a vector that has the magnetude of the shear force
and the direction of $\nabla u$.
The energy flux ${\mathbf J}$ and energy density ${\mathcal{U}}$
are given by
\beq
{\mathbf J}=-\mu_e\, v\,\nabla\, u,\;\;\;\;\;\;
{\mathcal{U}}=\frac{\rho}{2}\,v^2+\frac{\mu_e}{2}\,(\nabla u)^2,
\eeq
which are consistent with the results from the 3D
formula (See appendix A).

\subsubsection{${\mathbf J}$ and ${\mathcal{U}}$ for E-polarized wave}

For EM waves, we start with the consideration of the E-polarized
wave.
In this case the dynamical variables are ${\mathbf E}$ and
${\mathbf H}$ fields. Since $\nabla\cdot{\mathbf B}=0$, in terms
of the vector potential ${\mathbf A}$ we have
${\mathbf B}=\nabla\times {\mathbf A}$. Now we choose
\beq
{\mathbf A}=-\sqrt{4\pi c_0}\,\Phi\,\hat{\mathbf z}
\eeq
and
\beq
\alpha=\mu/c_0,\;\;\;\;\;\alpha c^2=c_0/\epsilon,
\eeq
then we have
\bqn
{\mathbf E}&=&-\frac{1}{c_0}\frac{\partial {\mathbf A}}{\partial t}
=\sqrt{\frac{4\pi}{c_0}}\,\varphi\,\hat{\mathbf z},\\
{\mathbf H}&=&\frac{1}{\mu}\nabla\times {\mathbf A}
=\sqrt{\frac{4\pi}{c_0}}\,{\mathbf Q}\times\hat{\mathbf z}.
\label{epeh}
\eqn
Using these relations then Eq.~(\ref{qphiwe}) indeed gives
the correct dynatimcal equations for E-polarized waves
(i.e., Eq.~(\ref{curle}) and (\ref{curlh})).
In addition, the textbook formulas for the Poynting vector
\beq
{\mathbf S}=\frac{c_0}{4\pi}{\mathbf E}\times{\mathbf H}
=\varphi\,\hat{\mathbf z}\times ({\mathbf Q}\times \hat{\mathbf z})
=\varphi\, {\mathbf Q}={\mathbf J},
\eeq
and the energy density
\beq
{\mathcal{W}}=\frac{1}{8\pi}(\epsilon E^2+\mu H^2)
=\frac{\alpha}{2}{\mathbf Q}^2+\frac{1}{2\alpha c^2}\varphi^2={\mathcal{U}}.
\eeq
give exactly the same results as those obtained from our formulas for energy flux
(Eq.~(\ref{flux})) and for energy density (Eq.~(\ref{ene})).

\subsubsection{${\mathbf J}$ and ${\mathcal{U}}$ for H-polarized waves}

In this case the dynamical variables are also ${\mathbf E}$ and
${\mathbf H}$ fields. Since we consider only the source free case,
we have $\nabla\cdot{\mathbf D}=0$. Thus we can introduce a vector
${\bf A}_m$ such that ${\mathbf D}=\nabla\times {\mathbf A}_m$. Here
${\mathbf A}_m$ represents another kind of vector potential,
used in deriving the ${\bf D}$ field.
Now we relate ${\mathbf A}_m$ to $\Phi$ via
\beq
{\mathbf A}_m=\sqrt{4\pi c_0}\,\Phi\,\hat{\mathbf z}
\eeq
and choose
\beq
\alpha=\epsilon/c_0,\;\;\;\;\;\alpha c^2=c_0/\mu,
\eeq
then we have
\bqn
{\mathbf H}&=&\frac{1}{c_0}\frac{\partial {\mathbf A}_m}{\partial t}
=\sqrt{\frac{4\pi}{c_0}}\,\varphi\,\hat{\mathbf z},\\
{\mathbf E}&=&\frac{1}{\epsilon}\nabla\times {\mathbf A}_m
=-\sqrt{\frac{4\pi}{c_0}}\,{\mathbf Q}\times\hat{\mathbf z}.
\eqn
Using these relations then Eq.~(\ref{qphiwe}) gives
the correct dynamical equations for H-polarized waves
(i.e., Eq.~(\ref{curle}) and (\ref{curlh})).
Furthermore, the Poynting vector
\beq
{\mathbf S}=-\frac{c_0}{4\pi}{\mathbf H}\times{\mathbf E}
=-\varphi\,\hat{\mathbf z}\times (-{\mathbf Q}\times \hat{\mathbf z})
=\varphi\, {\mathbf Q},
\eeq
together with the electromagnetic energy density
\beq
{\mathcal{W}}=\frac{1}{8\pi}(\epsilon E^2+\mu H^2)
=\frac{\alpha}{2}{\mathbf Q}^2+\frac{1}{2\alpha c^2}\varphi^2={\mathcal{U}}.
\eeq
are the same as those obtained from Eq.~(\ref{flux}) and Eq.~(\ref{ene}).

\subsection{Monochromatic waves}

Now we consider the monochromatic
wave. The auxiliary field $\Phi$ can be written as
\beq
\Phi=\frac{1}{2}
\left(\tilde{\Phi}+{\tilde{\Phi}}^*\right).
\eeq
Here ${\tilde{\Phi}}\propto e^{-i\omega t}$ satisfies
$\dot{\tilde{\Phi}}=-i\omega \tilde{\Phi}$.
$\tilde{\Phi}^*$ is the complex conjugate of $\tilde{\Phi}$.

The $\varphi$ and ${\mathbf Q}$ now become
\beq
\varphi=\frac{1}{2}\left(\tilde{\varphi}+\tilde{\varphi}^*\right),\;\;\;
{\mathbf Q}=\frac{1}{2}\left(\tilde{\mathbf Q}+\tilde{\mathbf Q}^*\right),
\eeq
with $\tilde{\varphi}$ and $\tilde{\mathbf Q}$ given by
\beq
\tilde{\varphi}=-i\omega \tilde{\Phi},\;\;\;\;
\tilde{\mathbf Q}=-\frac{1}{\alpha}\nabla\tilde{\Phi}.
\eeq
Since both $\tilde{\varphi}$ and $\tilde{\mathbf Q}$ contain a factor
$e^{-i\omega t}$, they can be factorized as
\beq
\tilde{\varphi}({\mathbf r},t)=\psi({\mathbf r})\,e^{-i\omega t},
\;\;
\tilde{\mathbf Q}({\mathbf r},t)=
\left(\frac{\nabla\psi({\mathbf r})}{i\omega \alpha({\mathbf r})}
\right)\,e^{-i\omega t},
\eeq
where $\psi$ is a time independent field.

The wave equation Eq.~(\ref{awe}) now becomes
\beq
\nabla\cdot\left(\frac{\nabla \psi({\bf r})}{\alpha({\bf r})}\right)+
\frac{\omega^2\psi({\bf r})}{\alpha({\bf r}) c^2({\bf r})}=0.\label{monowave}
\eeq

The time averaged energy flux and energy density are defined as
\bqn
\bar{\mathbf J}&\equiv&\frac{1}{T}\int^T_0 {\mathbf J(t)}\, dt\non\\
&=&
\frac{1}{2}\mbox{Re}\left(\tilde{\varphi}^* \,\tilde{\mathbf Q}\right)\non\\
&=&\frac{1}{2\alpha c}\mbox{Re}
\left( \frac{{\psi}^*\nabla{{\psi}}}{ik}\right)\label{flowj}
\eqn
and
\bqn
\bar{\mathcal{U}}&\equiv&\frac{1}{T}\int^T_0 {\mathcal{U}}(t)\,dt\non\\
&=&\frac{\alpha}{4}|\tilde{\mathbf Q}|^2
+\frac{1}{4\alpha c^2}|\tilde\varphi|^2\non\\
&=&\frac{1}{4\alpha c^2}\left(\,\left|\frac{\nabla {\psi}}{k}\right|^2
+\left|{\psi}\right|^2\right).\label{energyu}
\eqn
Here  $T=2\pi/\omega$ is the time period,
and $k=k({\mathbf r})=\omega/c({\mathbf r})$ is the wave number
in the media.

\section{Applications}

Now we turn to the applications of this unified theory. In the following
subsections we will discuss the use of the theory, only briefly, on two topics: the
wave scattering in a two dimensional system, and the band gap engineering problem.

\subsection{Scattering in a two dimensional system}

We discuss only a simple case. Suppose a (monochromatic) incident wave from
a source propagates to and be scattered by a circular cylinder located at the
origin.

Now we assume the radius of the cylinder is $a$.
The material parameter $\alpha({\bf r})$ inside and outside of the cylinder are given by constants
$\alpha_1$ and $\alpha$, respectively. Similarly, the wave speed are given by $c_1$ and $c$. We also define
$k_1=\omega/c_1$ and $k=\omega/c$ as the corresponding wave numbers, and denote the wave
inside and outside of the cylinder as $\psi_1({\bf r})$ and $\psi({\bf r})$.
According to Eq.(\ref{monowave}), they satisfy the Helmholtz equations,
\beq (\nabla^2 +k^2_1) \psi_1=0,\;\;\;\;(\nabla^2+k^2)\psi=0.\eeq

Since the Bessel and Hankel functions of all orders $\{J_n,H^{(1)}_n|
n=-\infty\rightarrow\infty\}$ forms a complete set of eigenfunctions, the solutions of
$\psi$ and $\psi_1$ can be written as the sums of them:
\bqn \psi(r,\theta)&=&\sum^{\infty}_{n=-\infty}[A_n H^{(1)}_n(kr)+B_n J_n(kr)]e^{in\theta},\non\\
\psi_1(r,\theta)&=&\sum^{\infty}_{n=-\infty}[C_n H^{(1)}_n(k_1r)+D_n
J_n(k_1r)]e^{in\theta}.\label{eq12}\eqn

To determine the coefficients, we have to find the boundary conditions.
According to Eq.(\ref{monowave}), and by using the divergence theorem, we get the boundary
conditions at $r=a$:
\beq \psi_1|_{r=a}=\psi|_{r=a},\;\;\;\;
\left.\frac{1}{\alpha_1}\frac{\partial\psi_1}{\partial r}\right|_{r=a}
=\left.\frac{1}{\alpha}\frac{\partial\psi}{\partial r}\right|_{r=a}.\label{boundary}\eeq

Furthermore, since the wave source is outside of the cylinder, the wave amplitude inside
the cylinder cannot goes to infinity, we therefore have $C_n=0$. Substitute Eq.(\ref{eq12}) and $C_n=0$ into
(\ref{boundary}), we have
\bqn
A_n H^{(1)}_n(ka)+B_n J_n(ka)&=&D_nJ_n(k_1a),\non\\
\frac{k}{\alpha}[A_n H^{(1)'}_n(ka)+B_n J'_n(ka)]&=&\frac{k_1}{\alpha_1}D_nJ'_n(k_1a).
\label{eq34}\eqn
These two equations lead to
\beq \frac{A_n}{B_n}=\frac{ghJ'_n(ka)J_n(ka/h)-J_n(ka)J'_n(ka/h)}{H^{(1)}_n(ka)J'_n(ka/h)
-ghH^{(1)'}_n(ka)J_n(ka/h)}\label{ab}\eeq and
\beq\frac{D_n}{B_n}
=\frac{2gh/(i\pi ka)}{H^{(1)}_n(ka)J'_n(ka/h)-ghH^{(1)'}_n(ka)J_n(ka/h)}.\label{db}
\eeq
Here we have defined $g\equiv\alpha_1/\alpha$ and $h\equiv c_1/c$. In deriving
Eq.(\ref{db}), we have used the identity
\[ J_n(x)H^{(1)'}_n(x)-J'_n(x)H^{(1)}(x)=\frac{2i}{\pi x}.\]

Suppose the incident wave is a plane wave:
\beq \psi_0({\bf r})=e^{ikx}=e^{ikr\cos\theta}
=\sum^{\infty}_{n=-\infty}i^nJ_n(kr)e^{in\theta},\label{inc}\eeq
then we have
\beq B_n=i^n.\eeq
Now the coefficients $A_n$ and $D_n$ can be obtained from Eq.(\ref{ab})
and (\ref{db}). The scattered wave is then given by
\beq \psi_{scat}({\bf r})\equiv\psi({\bf r})-\psi_0({\bf r})
=\sum^{\infty}_{n=-\infty}A_n H^{(1)}_n(kr)e^{in\theta}.\eeq

If the incident wave is not a plane wave, the only difference will be to give
a different set of $B_n$'s, and calcualte its corresponding $A_n$ and $D_n$'s from
Eq.(\ref{ab})and (\ref{db}).

\subsection{Band gap engineering}

Now we turn to the discussion about the photonic and phononic crystals.
In these {\it wave crystals} $\alpha({\bf r})$ and $c({\bf r})$ are periodic functions
of ${\bf r}$, i.e.,
\beq \alpha({\bf r}+{\bf R})=\alpha({\bf r}),\;\;\;\;\;c({\bf r})=c({\bf r}+{\bf R}).\eeq
Here
\beq {\bf R}=n_1{\bf a}_1+n_2{\bf a}_2\eeq
is an arbitrary translation vector, $n_1$ and $n_2$ are two integers, and
${\bf a}_1$ and ${\bf a}_2$ are the fundamental translation vectors \cite{Kittel}.
According to {\it Bloch's Theorem}, the eigenwave in wave crystal has
the form
\beq
\psi_{\bf K}({\bf r})=e^{i{\bf K}\cdot{\bf r}}\xi({\bf r}),\label{bloch}
\eeq
where ${\bf K}$ is the {\it Bloch wave vector}, and $\xi({\bf r})$ is a periodic
function, satisfying
\beq \xi({\bf r})=\xi({\bf r}+{\bf R}).
\eeq

The band structure of the wave crystal can be calculated in a unified manner from
Eq.(\ref{monowave}) by using the method that used in the second paper of
Ref.~\cite{Kush1}.
We will not go through the details of the band structure calculations but instead
give a brief discussion about the band gap engineering.

The main aim of the band gap engineering in the wave crystal study
is to create a large band gap.
Recently, Zhang {\it et.al} have developed a systematic method for
enlarging a photonic \cite{Zhang1} or phononic \cite{Zhang2} band gap.
Their methods are based on the variational principle.
Their formulas tell them how and where to vary the
material parameters can enlarge a band gap that already exists.
Here we will derive the unified version of their formulas
(See Eq.(5) and Eq.(7) of Ref.\onlinecite{Zhang1} and Eq.(2) of
Ref.\onlinecite{Zhang2}).

Suppose $\psi_{\bf K}({\bf r})\equiv\psi({\bf r})$ is
an eignfunction of the operator
\[ -\alpha({\bf r}) c^2({\bf r})\nabla\cdot\frac{1}{\alpha({\bf r})}\nabla,\]
and $\omega^2_{\bf K}\equiv\omega^2$ is the corresponding eigenvalue. That is, $\psi({\bf r})$ satisties
Eq. (\ref{monowave}). Multiplying
Eq.(\ref{monowave}) by $\psi^*$, we find
\beq \nabla\cdot\left(\frac{\psi^*\nabla\psi}{\alpha}\right)
-\frac{|\nabla\psi|^2}{\alpha}+\frac{\omega^2|\psi|^2}{\alpha c^2}=0.\label{three}\eeq
According to Eq.(\ref{bloch}), the quantity
\[ \frac{\psi^*\nabla\psi}{\alpha}
=\frac{i|\xi|^2{\bf K}+\xi^*\nabla\xi}{\alpha}\]
is a periodic vector field. Integrating Eq.(\ref{three}) over a unit cell, we get
\beq \int_{cell}\frac{|\nabla\psi({\bf r})|^2}{\alpha({\bf r})}d^2r
=\omega^2\int_{cell}\beta({\bf r})|\psi({\bf r})|^2d^2r.\label{twoenergy}
\eeq
Here we have defined
\beq \beta({\bf r})\equiv\frac{1}{\alpha({\bf r})c^2({\bf r})}.\eeq

Rewriting Eq.(\ref{monowave}) as
\beq \nabla\cdot \left(\frac{\nabla\psi}{\alpha}\right)=-\omega^2\beta\psi,\label{aphbt}\eeq
we can derive
\bqn &&\nabla\cdot\left[\delta\left(\frac{1}{\alpha}\right)\nabla\psi
+\frac{1}{\alpha}\nabla \delta \psi\right]\non\\&&=-\delta \omega^2 \beta \psi
-\omega^2 \left(\delta \beta \psi+\beta\delta
\psi\right).\label{vary1}\eqn
Note that in any case we {\it do not} vary the Bloch wave vector ${\bf K}$.

Multiplying Eq.(\ref{vary1}) by $\psi^*$ and integrating the resulting product
over one unit cell, we find
\bqn &&\int_{cell} \nabla\psi^*\cdot\left[\delta\left(\frac{1}{\alpha}\right)
\nabla\psi+\frac{1}{\alpha}\nabla\delta\psi\right]d^2r\non\\
&&=\delta\omega^2\int_{cell}\beta |\psi|^2d^2r\non\\
&&+\omega^2\int_{cell}
\left(\delta\beta|\psi|^2+\beta\psi^*\delta\psi\right)d^2r.\label{vary2}
\eqn
Multiplying the complex conjugation of Eq.(\ref{aphbt}) by $\delta\psi$,
we can also derive
\beq \int_{cell}\nabla\delta\psi\cdot\frac{\nabla\psi^*}{\alpha}d^2r
=\omega^2\int_{cell}\beta \psi^*\delta\psi.\label{vary3}d^2r\eeq
By substituting Eq.(\ref{vary3}) into Eq.(\ref{vary2}), and using
Eq.(\ref{twoenergy}), we finally obtain
\beq \frac{\delta\omega}{\omega}
=\frac{1}{2}
\left[\frac{\int_{cell}\delta\left(\frac{1}{\alpha}\right)|\nabla\psi|^2d^2r}
{\int_{cell}\frac{1}{\alpha}|\nabla\psi|^2d^2r}
-\frac{\int_{cell}\delta\beta|\psi|^2d^2r}{\int_{cell}\beta|\psi|^2d^2r}\right].\label{final}
\eeq
This is the unified version of the band gap engineering formula derived by Zhang
{\it et.al.}\cite{Zhang1,Zhang2}.

For acoustic wave we have $\psi=p$, $\alpha=\rho$ and $\beta=1/\alpha c^2$,
and hence
\[
\frac{\delta (\omega^2)}{\omega^2}=2\frac{\delta \omega}{\omega},\;\;\;
\delta \left(\frac{1}{\rho}\right)=-\frac{\delta\rho}{\rho^2},\;\;\;
\delta\beta=-\frac{\delta(\rho c^2)}{(\rho c^2)^2}.
\]
We therefore have from Eq.(\ref{final}) and Eq.(\ref{twoenergy}) the result
\bqn
\frac{\delta (\omega^2)}{\omega^2}&=&
\frac{\int_{cell}\frac{|p({\bf r})|^2\delta(\rho({\bf r}) c^2({\bf r}))}{(\rho({\bf r}) c^2({\bf r}))^2}d^2r}
{\int_{cell}\frac{|p({\bf r})|^2}{\rho({\bf r}) c^2({\bf r})}d^2r}
\non\\
&&-\frac{\int_{cell}\frac{|\nabla p({\bf r})|^2\delta\rho({\bf r})}{\rho^2({\bf r})}d^2r}
{\omega^2\int_{cell}\frac{|p({\bf r})|^2}{\rho({\bf r}) c^2({\bf r})}d^2r},
\eqn
which is just the Eq.(2) of Ref.\onlinecite{Zhang2}.
Results for the EM waves can also be obtained in a similar way.

We now give several remarks about Eq.(\ref{twoenergy}).

{\bf A}. In fact, Eq. (\ref{twoenergy}) is a consequence of the harmonic property of the wave field.
In the second line of Eq.(\ref{energyu}) we note that the energy density can be written
as the sum of two terms: $\alpha|\tilde{\bf Q}|^2/4=|\nabla\psi|^2/4\alpha\omega^2$ and
$|\varphi|^2/4\alpha c^2=|\psi|^2/4\alpha c^2$.
Let's call them the type I energy and the type II energy. Eq.(\ref{twoenergy}) means that
the time average of this two kinds of energies in one unit cell is equal. This is reasonable
because to form a harmonic wave the type I energy has to transform itself to type II completely
and then transform back in one time period $T/\omega$.

{\bf B}. From Eq.(\ref{twoenergy}), we have
\beq
\omega^2=\frac{\int_{cell}\frac{|\nabla\psi({\bf r})|^2}{\alpha({\bf r})}d^2r}
{\int_{cell}\beta({\bf r})|\psi({\bf r})|^2d^2r},
\eeq
which yields
\bqn \ln\omega^2&=&\ln{\int_{cell}\frac{|\nabla\psi({\bf r})|^2}{\alpha({\bf
r})}d^2r}\non\\
&-&\ln{\int_{cell}\beta({\bf r})|\psi({\bf r})|^2d^2r}.\label{log}\eqn
Taking the variation of Eq.(\ref{log}) and keeping the $|\psi|$ and $|\nabla\psi|$ terms
unchanged then gives us Eq.(\ref{final}). Why can we assume in this variational procedure
that $|\psi|$ is unchanged? The reason is that the change of $|\psi|$ caused by
the first order perturbation of $1/\alpha$ and $\beta$ is second order.

{\bf C}. Although in a unit cell the total type I energy
is equal to the total type II energy, however, their distributions are different.
Consider a binary system, i.e., the wave crystal that composed of two kinds of
homogeneous materials. A unit cell of such a wave crystal can be devided into
region {\it a} and region {\it b}, with material parameters $\alpha_a$, $\beta_a$ and $\alpha_b$, $\beta_b$,
respectively. According to Ref. \cite{Sig,CSK1,CSK2}, it is the impedance $\beta/\alpha$
that determine the size of the band gap. More explicitly, a wave crystal with large
$\beta_a/\alpha_a$ and small $\beta_b/\alpha_b$, or a small $\beta_a/\alpha_a$ and
large $\beta_b/\alpha_b$, will most probably have a large band gap.
This seems to imply that to have a large gap we have to design the
$\alpha$ and $\beta$ parameters such that the type I and type II
energies are efficiently separated in the unit cell.

\section{Summary}
In this paper, we have shown that for four kinds of classical
waves propagating in two-dimensional media, a unified treatment
can be constructed. These waves are the AC wave in fluid, the EL
shear (ELSH) wave in solid, and the E- and H-polarized EM waves in
nonabsorptive media. This unified theory helps us to find the
essentials of various wave phenomena and thus give us a more
complete and comprehensive understanding of these phenomena.

\section*{ACKNOWLEDGMENTS}
The authors would like to acknowledge helpful discussions with Dr.
Y. M. Kao, D. H. Lin, J. T. Lin, and M. C. Wu. This work was
supported by the National Science Council, Republic of China.

\appendix

\section{Conservation laws for elastic waves}

The equation of motion for elastic wave is of the form \beq \rho\,
\ddot{u}_i=\partial_j T_{ji}. \eeq Multiplying both side by
$\dot{u}_i$, we find \beq
\partial_t(\,\rho \dot{\mathbf u}^2/2\,)
=\partial_j\, (T_{ji}\,\dot{u}_i)-
T_{ji}\,\partial_t\,(\partial_j u_i),\label{rtu}
\eeq
where
\bqn
&&\hspace{-3mm}T_{ji}\,\partial_t\,(\partial_j u_i)\non\\
&=&\left[\,\lambda_e (\nabla\cdot {\mathbf u})\delta_{ij}
+\mu_e (\partial_i u_j+\partial_j u_i)\,\right]\partial_t\,(\partial_j u_i)\non\\
&=&\lambda_e\,(\nabla\cdot{\mathbf u})\partial_t\,(\nabla\cdot{\mathbf u})\non\\
&&+(\mu_e/2)\,(\partial_j u_i+\partial_i u_j)\partial_t (\partial_j u_i+\partial_i u_j)\non\\
&=&\lambda_e\,(\nabla\cdot{\mathbf u})\partial_t\,(\nabla\cdot{\mathbf u})
+2\mu_e\,E_{ij}\partial_t E_{ji}\non\\
&=&\partial_t\,\left[\lambda_e\,(\nabla\cdot{\mathbf u})^2/2
+ \mu_e\,\mbox{Tr}\,({\mathbf E}^2) \right].\label{tu}
\eqn
In deriving (\ref{tu}) we have defined a symmetric matrix ${\mathbf E}$:
\beq
E_{ij}=\frac{1}{2}(\partial_i u_j+\partial_j u_i).
\eeq
Substituting (\ref{tu}) into (\ref{rtu}) one finds the conservation law
\beq
\nabla\cdot\,({\mathbf T}\cdot\dot{\mathbf u})=\frac{\partial}{\partial t}
\left[\frac{\rho}{2}\, \dot{\mathbf u}^2+\frac{\lambda_e}{2}\,(\nabla\cdot{\mathbf u})^2
+ \mu_e\,\mbox{Tr}\,({\mathbf E}^2) \right].
\eeq
Thus the energy flux ${\mathbf J}$ and energy density ${\cal U}$
are
\beq
{\mathbf J}=-{\mathbf T}\cdot\dot{\mathbf u},\;\;
{\cal U}=\frac{\rho}{2}\, \dot{\mathbf u}^2+\frac{\lambda_e}{2}\,(\nabla\cdot{\mathbf u})^2
+ \mu_e\,\mbox{Tr}\,({\mathbf E}^2).
\eeq

For the special case that acoustic waves propagating in fluid, we have
$\mu_e=0$. Using the relation $p=-\lambda\nabla\cdot{\mathbf u}$,
we find
\beq
{\mathbf T}=-p\,{\mathbf I},
\eeq
where ${\mathbf I}$ is the unit matrix.

Denoting
$\dot{\mathbf u}$ as ${\mathbf v}$ and remembering that
$\lambda_e=\rho c^2$ then one finds
\beq
{\mathbf J}=p\,{\mathbf v},\;\;\;\;\;
{\cal U}=\frac{\rho}{2}\,{\mathbf v}^2+\frac{1}{2\rho c^2}\,p^2.
\eeq

Another special case is the elastic SH wave.
In this case ${\mathbf u}=u\hat{\mathbf z}$. Remember that
$\partial u/\partial z=0$, therefore $\nabla\cdot{\mathbf u}=0$ and we have
\bqn
{\mathbf T}&=&2\mu_e {\mathbf E},\\
{\mathbf E}&=&\frac{1}{2}\left(\bary{ccc}0&0&\frac{\partial u}{\partial x}\\
0&0&\frac{\partial u}{\partial y}\\
\frac{\partial u}{\partial x}&\frac{\partial u}{\partial y}&0\eary\right)\non\\
\Longrightarrow\; \mbox{Tr}({\mathbf E}^2)
&=&\frac{1}{2}(\nabla u)^2.
\eqn

The energy flow and energy density are thus given by
\beq
{\mathbf J}=-\mu_e\,v\nabla\,u,\;\;\;\;
{\cal U}
=\frac{\rho}{2}\,{\mathbf v}^2+\frac{\mu_e}{2\,}\,(\nabla u)^2.
\eeq
Here ${\mathbf v}=\dot{\mathbf u}=\dot{u}\,\hat{\mathbf z}$ is the vibration
vecocity of the media.

\end{document}